\documentclass{natureJLH}
\usepackage[english]{babel}
\usepackage{lineno, blindtext}
\usepackage{epsfig}
\usepackage{subcaption} 
\usepackage{ae, aecompl}
\usepackage{helvet}
\usepackage{epsfig}
\usepackage{dcolumn}
\usepackage{color}
\usepackage[LGR, T1]{fontenc}
\usepackage[latin9]{inputenc}
\usepackage{amssymb}
\usepackage{amsmath}
\usepackage{mwe, tikz}
\usepackage{esint}
\usepackage{longtable}
\usepackage{url}
\usepackage{etoolbox}
\usepackage{multirow}
\usepackage{tcolorbox}

\newcommand\apj{Astrophys. J.,}
\newcommand\apjl{Astrophys. J. Lett.,}

\newcommand\araa{Annu. Rev. Astron. Astrophys.,}
\newcommand\mnras{Mon. Not. R. Astron. Soc.,}
\newcommand\nat{Nature,}
\newcommand\physrep{Phys. Rep.,}

\newcommand\prd{Phys. Rev. D}
\newcommand\aapr{Astron. Astrophys. Rev.}
\newcommand{\aap}{Astron. Astrophys.}

\newcommand{\pasj}{Pub. Astron. Soc. Japan}

\usepackage{xpatch}

\title{The Physical Mechanisms of Fast Radio Bursts}

\author{Bing Zhang}

\begin{document}

\maketitle

\begin{affiliations}
\item Department of Physics and Astronomy, University of Nevada, Las Vegas, NV 89154, USA; 
Email: zhang@physics.unlv.edu, https://orcid.org/0000-0002-9725-2524. 
\end{affiliations}


\begin{abstract}
Fast radio bursts are mysterious millisecond-duration transients prevalent in the radio sky. Rapid accumulation of data in recent years has facilitated an understanding of the underlying physical mechanisms of these events. Knowledge gained from the neighboring fields of gamma-ray bursts and radio pulsars also offered insight. Here I review developments in this fast-moving field.Two generic categories of radiation model invoking either magnetospheres of compact objects (neutron stars or black holes) or relativistic shocks launched from such objects have been much debated. The recent detection of a Galactic fast radio burst in association with a soft gamma-ray repeater suggests that magnetar engines can produce at least some, and probably all, fast radio bursts. Other engines that could produce fast radio bursts are not required, but are also not impossible.
\end{abstract}

On 24 July, 2001, a bright, dispersed radio pulse reached the Parkes 64-m telescope in Australia. The recorded signal remained unnoticed until Duncan Lorimer and his collaborators discovered it (now known as the fast radio burst FRB 010724) several years later and published the results in Science\cite{lorimer07}. For some time, some astronomers were not convinced that this so-called ``Lorimer burst'' was a genuine type of astrophysical event, until four more events were reported in 2013 (ref. \cite{thornton13}). The field then enjoyed a rapid boost from both observational and theoretical fronts. With the identification of the microwave-oven-origin of some artificial bursts known as ``perytons'', the astrophysical origin of the bona fide bursts was finally established\cite{petroff15c}. This was followed by the detection of the first repeating source, known as FRB 121102 (ref. \cite{spitler16}) and its localization in a dwarf star forming galaxy at redshift $z = 0.19$(refs. \cite{chatterjee17,marcote17,tendulkar17}), establishing the extragalactic/cosmological origin (in contrast to the Galactic origin\cite{loeb14}) of these events. Numerous theoretical models have been proposed to interpret these mysterious events\cite{platts19}. These have been driven by observations and ground-breaking results from numerous radio observatories: Parkes, Arecibo, Green Bank Telescope, the Australian Square Kilometre Array Pathfinder (ASKAP), the Canadian Hydrogen Intensity Mapping Experiment (CHIME), the Five-hundred-meter Aperture Spherical radio Telescope (FAST), the Survey for Transient Astronomical Radio Emission 2  (STARE2), to name a few. Approximately every six months during the past seven years, our knowledge about FRBs has undergone a quantum leap. The prosperity of the field is also marked by the steady growth of publications and citations, already exceeding those in the early years in the field of gamma-ray bursts (GRBs)\cite{kulkarni18}. On 28 April, 2020, one FRB-like event was detected by CHIME\cite{CHIME-SGR} and STARE2\cite{STARE2-SGR} in association with a hard X-ray burst from a Galactic magnetar, the soft gamma-ray repeater SGR 1935+2154 during its active phase\cite{HXMT-SGR,Konus-SGR,Integral-SGR,AGILE-SGR}. This discovery established magnetars as a source that can produce FRBs.

The FRB science has been reviewed several times, focusing mostly on the rapid observational progress\cite{petroff19,cordes19,lorimer18}, but sometimes on surveys of many theoretical models\cite{katz18,popov18,platts19}. This article summarizes major observational facts, critically reviews the theoretical models on radiation mechanisms and FRB sources, and discusses open questions in the field.

\section{Observational facts.}

    {\em Duration.}
    The typical observed FRB duration (also known as width $w$) is some milliseconds. This defines a characteristic length scale of the engine that powers FRBs, i.e. $l_{\rm eng} \lesssim c w = (3\times 10^7) \ {\rm cm} \ (w / {\rm ms})$, where $c$ is the speed of light. This immediately points towards the most compact objects in the universe, that is, a neutron star or a stellar-mass black hole. 

{\em Repetition.}
     Since FRB 121102 (ref. \cite{spitler16}), more than 20 FRBs have been reported to also repeat\cite{chime-2ndrepeater,chime-repeaters,kumar19,luo20}. It is tempting to speculate that the majority or even all FRB sources are repeaters\cite{ravi19b,lu20b} with the apparently non-repeating FRBs being the sources with longer quiescent periods or with bursts too faint to be detected, even though the existence of genuinely non-repeating FRBs cannot be ruled out with the current data\cite{petroff15b,palaniswamy18,caleb19}. 

    {\em Periodicity.} For repeating sources,  a period of the order of milliseconds to seconds would indicate a neutron-star origin for FRB sources, but a dedicated search for a period in this range\cite{zhangy18} has been unsuccessful.
Surprisingly, a longterm period of $16.35\pm 0.15$ days with an approximately 5-day active window has been reported for the CHIME repeating source FRB 180916.J0158+65\cite{chime-periodic}. A plausible period of about $157$ days has also been suggested for FRB 121102 (ref.\cite{rajwade20}). Such a longterm periodicity can be accommodated within the framework of binary\cite{ioka20,lyutikov20,dai20} or precession\cite{levin20,zanazzi20,yangzou20} models. 

     {\em Pulse morphology.}
FRB lightcurves show diverse shapes. A fraction of FRBs have single pulses, some accompanied by a well defined decaying tail consistent with scattering in a cold plasma ($w \propto \nu^{-4}$ or $w \propto \nu^{-4.4}$, where $\nu$ is the observing frequency), which carries the information about the turbulent properties of interstellar medium in the host galaxy or the plasma in the vicinity of the bursting source\cite{luan14,cordes16b,xu16}. Some FRBs show complex pulse structures. Some (mostly from repeating FRB sources) show a clear sub-pulse frequency down-drifting signature\cite{hessels19,chime-2ndrepeater,chime-repeaters}, that is, the sub-pulses with progressively lower frequencies arrive at progressively later times. 

     {\em Dispersion measure.} The universe is not empty but filled with interstellar and intergalactic gas and plasma. Propagation of radio waves is affected by the existence of free electrons along the way, with lower-frequency waves traveling more slowly than high-frequency waves. The degree of such delay is defined by the parameter called dispersion measure (DM), which carries the physical meaning    $ {\rm DM} = \int_0^{D_z} \frac{n_e(l)}{1+z(l)} \ d l$, that is, the integral of free (ionized) electron number density $n_e$ along the line of sight from the source to the observer, corrected by the cosmic expansion factor $(1+z)$ (where $z$ is the redshift, $l$ is length, and $D_z$ is the line-of-sight comoving distance of the source).  The observed DM values of FRBs are usually in great excess of the values allowed by the Milky Way Galaxy, suggesting that the FRB sources are at extragalactic or cosmological distances. In units of  $\rm pc \cdot cm^{-3}$, the measured DM values of the published FRBs\cite{FRBcat} range from about $100$ to $2600$, with a typical value around $300-400$. 
   
     {\em Redshift.} 
      Sub-arcsecond localizations of an increasing sample of FRBs have been accomplished using telescope arrays such as Karl G. Jansky Very Large Array, Deep Synoptic Array, and ASKAP. These localizations allow identifications of the host galaxies, and hence, the redshifts of a sample of FRBs, both repeating ones and apparently non-repeating ones: FRB 121102 at $z=0.19$ (ref. \cite{tendulkar17}), FRB 180924 at $z=0.32$ (ref. \cite{bannister19}), FRB 190523 at $z=0.66$ (ref. \cite{ravi19}), FRB 180916 at $z=0.03$ (ref. \cite{marcote20}), FRB 181112 at $z=0.48$ (ref. \cite{prochaska19}),  and FRBs 190102, 190608, 190611, 190711 from $z=0.29, 0.12, 0.38, 0.52$, respectively\cite{macquart20}. The redshifts of these events are in general agreement with the excess DM measured for the respective FRBs within the framework of the standard cosmology, with the baryonic mass density ($\Omega_b$) and the fraction of baryons in the intergalactic medium ($f_{\rm IGM}$) consistent with previous measurements\cite{lizx20,macquart20}.
     
     {\em Luminosity and energetics.} The establishment of the distance scale of FRBs allows one to calculate the luminosity and energy of the FRBs.  The isotropic equivalent peak luminosities of known FRBs vary from\cite{STARE2-SGR,zhang18a,ravi19} about $10^{38} \ {\rm erg \ s^{-1}}$ to a few $10^{46} \ {\rm erg \ s^{-1}}$. The corresponding isotropic energies vary from a few $10^{35}$ erg to a few $10^{43}$ erg. The luminosity is extremely high by radio pulsar standards, but is minuscule by GRB standards. The true energetics of FRBs should be reduced by a beaming factor $f_b = {\rm max} (\Delta\Omega/4\pi, 1/4\gamma^2) \leq 1$, where $\Delta\Omega$ is the solid angle of the geometric beam, and $\gamma$ is the Lorentz factor of the FRB emitter ($1/\gamma$ is the half kinetic beaming angle for an FRB emitter traveling close to the speed of light). A successful FRB engine should at least generate a luminosity and an energy of the order of $f_b L_p$ and $f_b E$, respectively. Observationally, the majority of hard X-ray bursts from SGR 1935+2154 were not associated with FRBs\cite{lin20}. This is consistent with the FRB emitters (at least those produced by magnetars) being narrowly beamed. 
    
    {\em Brightness temperature.} The combination of high luminosity and short variability timescale of an FRB defines an extremely high brightness temperature $T_b \simeq \frac{{\cal S}_{\nu,p} D_{\rm A}^2}{2 \pi k (\nu \Delta t)^2} = (1.2\times 10^{36} \ {\rm K} ) \left( \frac{D_{\rm A}}{10^{28} \ {\rm cm}} \right)^2 \frac{{\cal S}_{\nu,p}}{\rm Jy} \left(\frac{ \nu}{\rm GHz}\right)^{-2} \left(\frac{\Delta t}{\rm ms}\right)^{-2}$, where $k$ is Boltzmann constant, $\Delta t$ is the variability time scale ($\Delta t=w$ for single-pulse bursts),  and $D_{\rm A} =D_z/(1+z) $ is the angular distance of the source. The gigantic $T_b$ ($\sim 10^{36}$ K for nominal FRB parameters) is much greater than approximately $10^{12}$ K, the maximum $T_b$ achievable for incoherent radiation\cite{kellermann69}. This demands that the radiation mechanism for FRB emission must be ``coherent'', that is, the radiation by relativistic electrons is not only not absorbed, but is also greatly enhanced with respect to the total expected emission if electrons radiate independently (or incoherently). Before the discovery of FRBs, radio pulsars have been the known sources that produce extremely high $T_b$ values (typically about $(10^{25}-10^{30})$ K). FRBs further push the limit of the degree of coherent radiation in the universe.  

    {\em Typical frequency and  spectrum.} FRBs have been detected from 300 MHz (ref. \cite{chawla20}) to at least 8 GHz (ref. \cite{gajjar18}). Non-detection at higher frequencies is probably due to the limited sensitivity\cite{law17}, but non-detection at lower frequencies, especially with LOFAR at 145 MHz, probably suggests an intrinsic hardening of the spectrum at low frequencies\cite{karastergiou15}. 
    The spectral indices of bursts from the repeater FRB 121102 range from -10 to +14 (ref. \cite{spitler16}) when a simple power law model is used, which may hint a narrow spectrum for FRBs. A narrow spectrum is also suggested by the relative flux (or flux limit) of the two pulses of the Galactic FRB 200428 as detected by CHIME and STARE2 in two different energy bands\cite{CHIME-SGR,STARE2-SGR}. 

    {\em Polarization properties.} 
    Most FRBs have strong linear polarization typically with a degree of polarization above 50\%, sometimes near 100\%\cite{michilli18,luo20,cho20,day20}. 
    These properties are in general similar to those of radio pulsars but with noticeable differences.
    A good fraction of radio pulsars are characterized by the ``S'' or inverse ``S'' shape patterns\cite{lorimer12}. Some are consistent with the so-called ``rotation vector'' model\cite{rv69}. 
    Swings of the PA across pulses are indeed observed in some FRBs, but may not follow the simplest rotating vector model\cite{cho20,luo20}. 
    Some FRBs show a constant polarization angle (PA) across each pulse\cite{michilli18}. 

    {\em Rotation measure.} If the medium through which radio waves propagate through is magnetized, the linear polarization angle would undergo a frequency-dependent variation known as ``Faraday rotation''. The degree of rotation is given by the so-called rotation measure  (RM), 
    $   (-0.81 \ {\rm rad \ m^{-2}}) \ \int_0^{D_z}  \frac{ [B_\parallel(l) / {\rm mG}] n_e(l)}{[1+z(l)]^2} dl$, where $B_\parallel(l)$ is the $l$-dependent magnetic field strength along the line of sight (in units of milliGauss). The measured absolute values of the RM for FRBs range from nearly zero\cite{ravi16} to around $10^5 \ {\rm rad \ m^{-2}}$ in the case of FRB 121102 (ref. \cite{michilli18}).
     The RM values of some FRBs show either a secular\cite{michilli18} or short-term\cite{luo20} evolutionary trend.

    {\em Steady radio source.} In the radio band, the first repeater FRB 121102 was found to be associated with a steady radio source\cite{chatterjee17}. The flux varies with time with a mean flux density of about $190 \mu$Jy. 
    Such a bright steady source is rare among FRBs. It may be related to the special environment (for example, it is in a young magnetar wind nebula or near a supermassive black hole) and extremely large RM of FRB 121102 (refs. \cite{michilli18,magalit18,yang20b}). 
    
     {\em Multi-wavelength counterparts.} Multi-wavelength follow-up observations did not reveal any ``afterglow''-like emission\cite{petroff15}, consistent with theoretical predictions\cite{yi14}. Searches for FRBs in association with GRBs during the prompt phase have not delivered confirmed results\cite{bannister12,delaunay16,cunningham19}. Neither have searches for young-magnetar-powered FRBs\cite{metzger17} from the remnants of long GRBs or superluminous supernovae\cite{law19,men19} (but see Ref. \cite{wangxg20}). 
    So far, the only confirmed counterpart of an FRB is a hard X-ray/soft $\gamma$-ray burst\cite{HXMT-SGR,Konus-SGR,AGILE-SGR,Integral-SGR} from the Galactic magnetar SGR 1935+2154 that was spatially and temporally associated with FRB 200428 (refs. \cite{CHIME-SGR,STARE2-SGR}). There are two hard spikes in the X-ray lightcurve that coincide with the two pulses of the FRB detected by CHIME, which unambiguously established magnetars as the source of at least some FRBs\cite{HXMT-SGR,Konus-SGR}. The non-detection of any FRB by FAST during the epochs of 29 other X-ray bursts suggests that the FRB-SGR burst associations are rather rare\cite{lin20}.

    {\em Spatial distribution.} The directional distribution of FRBs is consistent with being isotropic\cite{bhandari18}, similar to GRBs.  This is consistent with the cosmological origin of FRBs.

    {\em Luminosity/energy function.}  The global luminosity/energy function of FRBs (for both repeating and non-repeating ones) are consistent with a power-law function with index\cite{luo18,lu19b,luo20b} approximately $-1.8$ ($dN/dL \propto L^{-1.8}$ or $dN/dE \propto E^{-1.8}$) with a possible exponential cutoff at the high end\cite{luo18,luo20b}. At the low end, The function extends all the way down to the luminosity of the Galactic FRB 200428 (about $10^{38} \ {\rm erg \ s^{-1}}$)\cite{lu20}.  
    
    {\em Event rate density.} The observed FRB event rate is roughly several thousands over the whole sky per day above a threshold pf $1 \ {\rm Jy \ ms}$ fluence\cite{petroff19,cordes19}. Overall, the event-rate density of FRBs reaches\cite{luo20b} about $3.5\times 10^4 \ {\rm Gpc^{-3} \ {yr}^{-1}}$ above approximately $10^{42} \ {\rm erg \ s^{-1}}$ (and is even larger at lower luminosities), already exceeding the event-rate densities of most catastrophic transient events in the universe (including supernovae). The true rate should be increased by a factor of $1/f_b$. This disfavors catastrophic events as the sources of the majority of FRBs\cite{ravi19b,luo20b}. 
    
    {\em Host galaxies.} The host galaxy of FRB 121102 is analogous to those of long GRBs and superluminous supernovae\cite{tendulkar17,nicholl17}, but it is atypical.  Most other FRB hosts are massive galaxies with moderate to low star formation rates, and the locations of FRBs within the galaxies are not necessarily in the brightest region, sometimes with a large offset from the galactic centre\cite{bannister19,ravi19,bhandari20,marcote20}. The host galaxy types and the local environment of most FRBs are not consistent with those of long GRBs and superluminous supernovae, but rather resemble those of normal (Type Ia, Type II) supernovae and short GRBs\cite{lizhang20}. The data are consistent with the magnetar engine scenario if both normal magnetars (those observed in the Galaxy) and magnetars born of either GRBs or superluminous supernovae can produce FRBs\cite{lizhang20}.
    The probability that a good fraction of FRBs are related to neutron star - neutron star (NS-NS) mergers is also possible. Since the one-off FRB models invoking NS-NS mergers\cite{totani13,zhang14,wang16} cannot meet the event rate density requirement, repeating FRB models invoking  post-merger long-lived magnetars\cite{margalit19,wang20} or pre-merger magnetospheric interaction\cite{zhang20} are plausible FRB models invoking NS-NS mergers.

\section{Insight from neighboring fields.}

The sparse data of FRBs do not allow us to firmly identify the sources and mechanisms that produce them. Insight from neighboring fields can offer help. The two closest fields to the FRB field are that of GRBs\cite{zhang18}, which are more energetic cosmological bursting sources of $\gamma$-rays, and that of radio pulsars\cite{lorimer12}, which emit coherent radio emission with high brightness temperatures only dwarfed by FRBs. 

\begin{figure}
\begin{tabular}{c}
\includegraphics[width=1.0\textwidth]{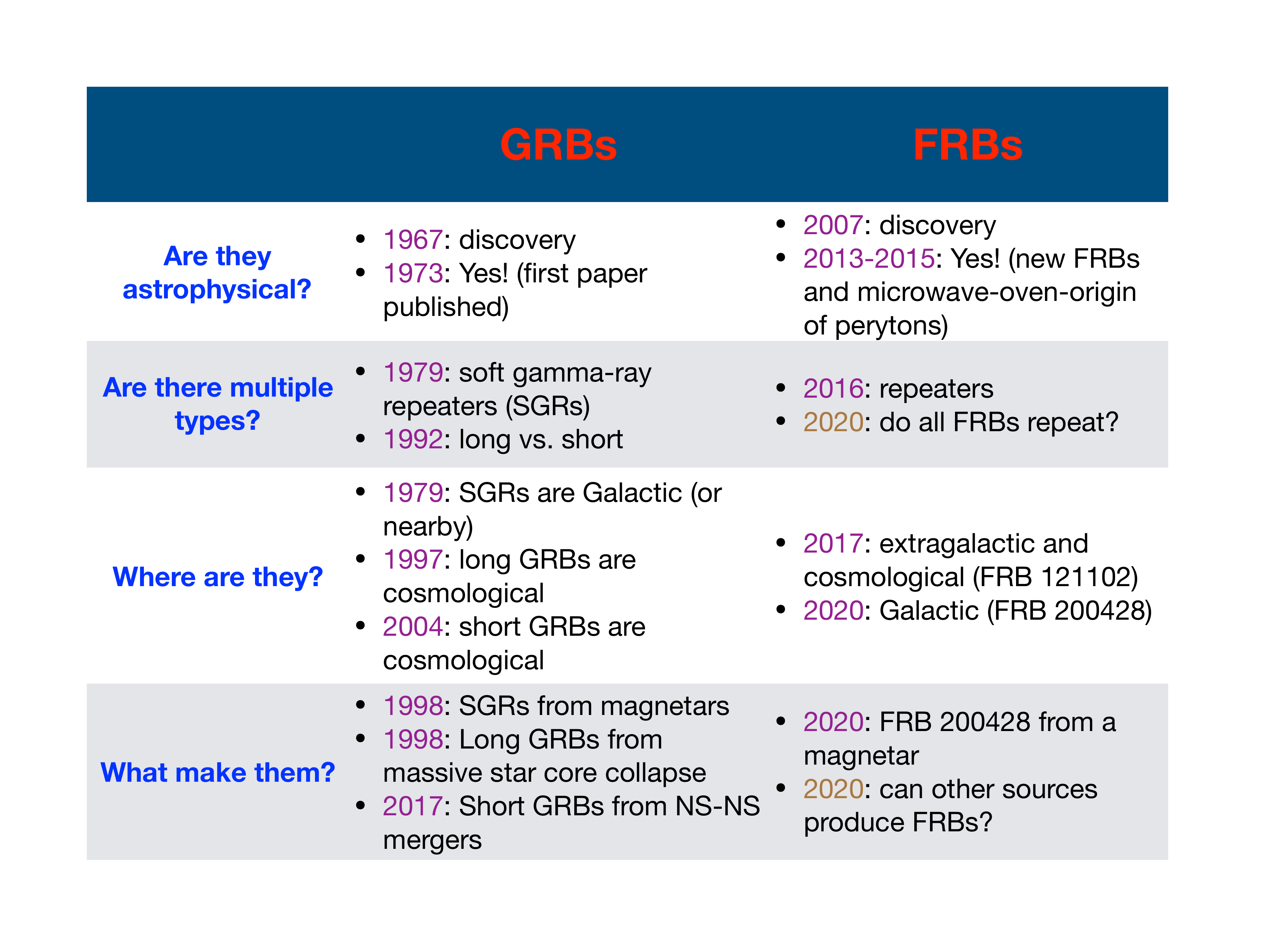}  \\
\end{tabular}
\caption{\bf A historical comparison between the GRB and FRB fields.}
\label{fig:history}
\end{figure}

The first lesson learned from the GRB field is its history. The GRB field went through several major stages characterized by seeking the definite answers to the following questions (Fig. \ref{fig:history}). \\
1. Are they astrophysical?  \\
2. Are there multiple types? \\
3. Where are they? \\
4. What made them?  \\
The FRB field is following essentially the same path, but at an accelerated pace. The first question in both fields took several years to answer. For the second question, there is still no clarity regarding whether there exist multiple types of FRB. Even though some FRBs repeat while others do not, it is essentially impossible to rule out the ansatz that {\em all FRBs repeat}. Interestingly, the early years of the GRB study were also complicated by repeaters, which for a time made at least some workers believe that all GRBs should eventually repeat. The confusion was removed after repeaters (Galactic magnetars) were separated from cosmological, catastrophic GRBs on the basis of their very different distance scales, and hence energy budgets. The FRB field is facing a similar but more difficult challenge. Repeaters seem to be prevailing and they are already at cosmological distances! 
Some observational properties (for example, longer durations and down-drifting sub-pulses) have been claimed as possible criteria with which to identify repeaters\cite{chime-repeaters}. However, because some repeating bursts do not share these properties, it is hard to draw the conclusion that those apparent non-repeaters that do not show these properties must not repeat.  
The answer to the third question for FRBs came much sooner than GRBs thanks to the superb localization capability offered by very long baseline interferometry. The result was unsurprising: the nominal FRBs (either repeating or apparently non-repeating) are at cosmological distances. Because of their very high event rate density, FRBs from the Milky Way or nearby galaxies should be detectable within human timescales, and indeed a low-luminosity FRB 200428 was detected during an X-ray burst from the Galactic magnetar SGR 1935+2154 (refs. \cite{CHIME-SGR,STARE2-SGR}). This detection also brought the answer to the fourth question: magnetars can make FRBs, as has long been speculated since the discovery of FRBs\cite{popov10,kulkarni14,lyubarsky14,katz16,metzger17,beloborodov17,kumar17,yangzhang18,wadiasingh20}.

There are two more ingredients that have been transplanted from the GRB field to the FRB field. First, owing to their similarly mysterious nature, many of the 120 or so early models for GRBs\cite{nemiroff94} were reinvented for FRBs\cite{platts19}. However, Malvin Ruderman said of GRBs  in 1975, ``the only feature that all but one (and perhaps all) of the very many proposed models have in common is that they are not the explanation of GRBs'', and we can say the same for FRBs today, with the small modification of ``all but one'' to ``all but a few'' to remain optimistic. Second, the basic relativistic synchrotron shock (both internal shocks and external shocks) models for GRBs have been found to be useful in the FRB field with the additional ingredient of synchrotron masers\cite{lyubarsky14,waxman17,beloborodov17,plotnikov19,metzger19,beloborodov20}, which may offer an interpretation to the high brightness temperature of FRB emission.

On the other hand, lessons can be learned from the radio pulsar field regarding coherent radiation mechanisms.  However, after more than 50 years of study, the radiation mechanism that is operating in pulsars is still not identified\cite{melrose17}. It is known that pulsar radio emission is broad band and much brighter than radio emission of the Sun and Jupiter so that the simple version of the coherent mechanisms applied to solar system objects do not apply\cite{melrose17}. Strong magnetic fields and acceleration of relativistic particles likely play an essential part in generating the bright coherent emission from pulsars. Broadly speaking, two main categories of models have been discussed in the literature. One category is the ``antenna'' models, which invoke physical clustering of particles in bunches with the size smaller than the radio wavelength, so that particles can radiate in phase similar to those electrons in human-made antennas. The other category is 
broadly termed ``maser'' models. It invokes either run-away growth of plasma oscillation modes that eventually escape as electromagnetic waves (also called ``plasma maser'') or ``negative'' absorption of photons in the emission region. 
In any case, the amplitude of radiation grows rapidly during propagation and eventually reaches a very high brightness temperature. 

Phenomenologically, there are two well known distinct radio emission sites in pulsars, which are manifested in the famous young Crab pulsar. Its radio pulses include three pulse components: Two of them align well with the high energy ($\gamma$-ray, X-ray, and optical) emission pulses and are separated by about 180$^{\rm o}$ in phase. Modeling suggests that they originate from the same emission region as high-energy photons, likely in the outer magnetosphere or even outside the light cylinder\cite{harding16}. The third pulse component is the ``precursor'' emission, which leads one of the widely-separated pulse components. Geometric modeling suggests that this component originates from the open field line region not far away from the surface of the neutron star. In fact, radio emission from most of the pulsars is consistent with this inner-magnetospheric pulse component\cite{rankin93,ruderman75}. Galactic magnetars normally do not emit radio pulses, but can become radio pulsars after bursting activities. Radio emission from magnetars show somewhat different spectral and temporal properties from radio pulsars\cite{camilo07}. The same pulsar emission mechanism may well be at play, but a completely different mechanism is not impossible. 

\section{Radiation mechanisms.}

\begin{figure}
\begin{tabular}{c}
\includegraphics[keepaspectratio, clip, width=1.0\textwidth]{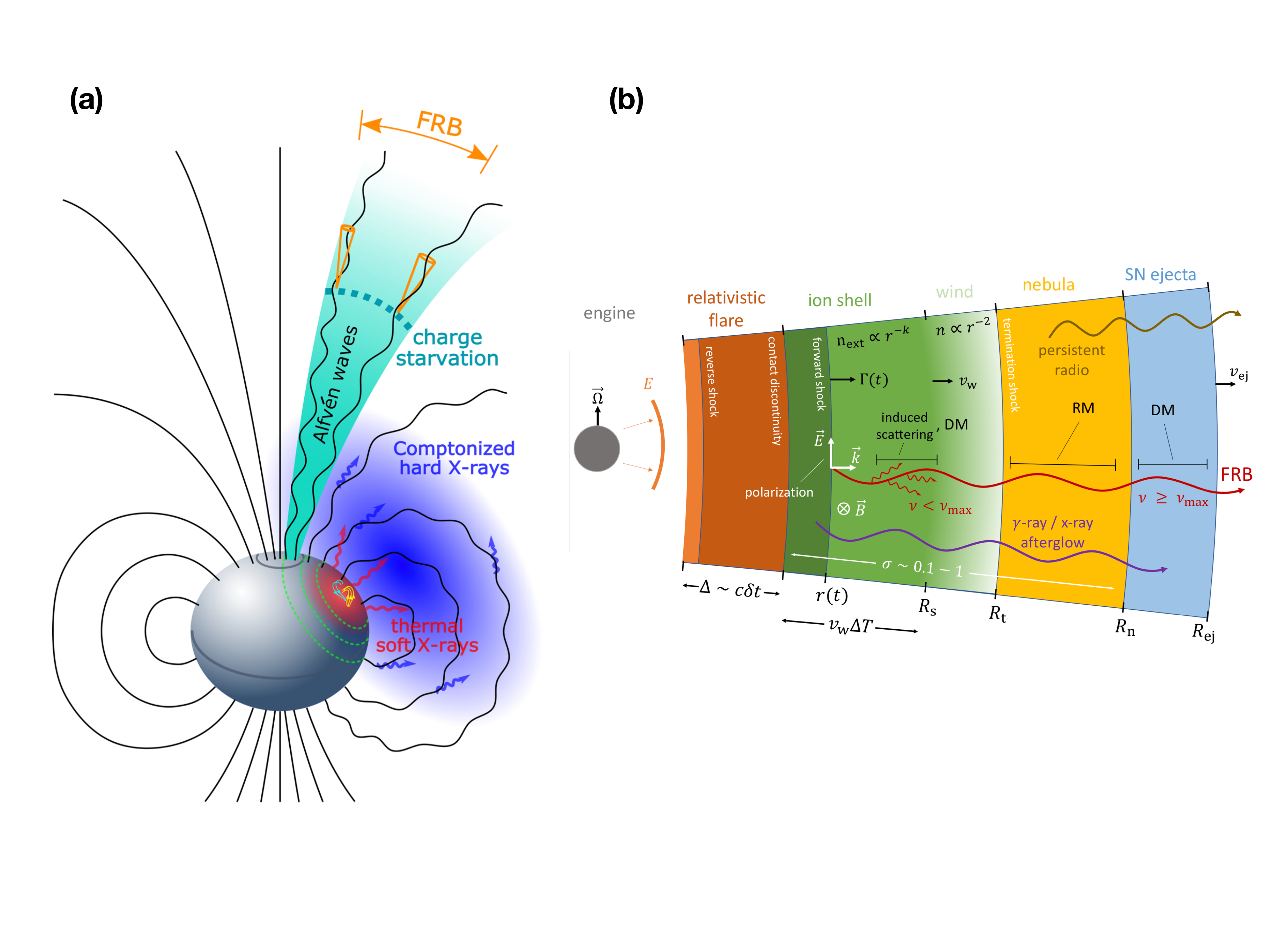} \\
\end{tabular}
\caption{{\bf Cartoon pictures of the two general types of FRB radiation models. a,} Pulsar-like models that invoke the magnetosphere of a compact object\cite{lu20}. {\bf b,} GRB-like models that invoke relativistic shocks launched from a compact object\cite{metzger19}. Magnetars can be the common source for both models.}
\label{fig:cartoons}
\end{figure}

A competitive FRB model should include two parts: (1) an energy source model that can account for the global properties of FRBs (energetics, event rate density, redshift distribution, host galaxy properties,  location in the host, as well as the properties of the immediate environment -- such as steady radio source, host DM, RM, and so on) -- and (2) a radiation mechanism model to explain the coherent radio emission with observed properties (brightness temperature, duration, temporal and spectral features, as well as polarization features). Whereas the first part of the model allows for a wide range of speculations (there have been more than fifty of those), the second part permits only limited possibilities, which can be discussed generically regardless of the source models. 
Very generally, one can group the FRB coherent radio emission models into two categories (Figure \ref{fig:cartoons}): those invoking the magnetospheres of a compact object (probably a neutron star, but a black hole or an ``effective'' pulsar formed by compact binary coalescence\cite{zhang16a,levin18} is also possible) and those invoking relativistic shocks far outside the magnetospheres. We may call the former models ``pulsar-like'' and the latter models ``GRB-like''. 
Regardless of the type of the model, the typical frequency of a coherent mechanism may be related to one of the following characteristic frequencies: (1) the plasma angular frequency $\omega_p$, its Doppler boost ${\cal D} \omega'_p$ (where $\cal D$ is the Doppler factor and $\omega'_p$ is the comoving-frame plasma frequency) or related frequencies such as the Razin frequency\cite{waxman17,plotnikov19,long18};  (2) the cyclotron angular frequency $\omega_B$ or its Doppler boost ${\cal D} \omega'_B$ (refs. \cite{lyubarsky14,beloborodov17,metzger19,beloborodov20}); and (3) the characteristic angular frequency of a certain radiation mechanism (for example, curvature or synchrotron radiation) of relativistic particles\cite{katz14,kumar17,lu18,yangzhang18,ghisellini17}. Regardless of the radiation mechanism, in order for radio emission with an extremely high $T_b$ to reach Earth, the waves must also overcome several absorption effects including plasma damping\cite{lu19}, induced Compton scattering\cite{lyubarsky08,murase16,lu18,kumar20b},  ``free-free'' absorption\cite{murase16,piro16,metzger17,yangzhang17}, and synchrotron absorption by the surrounding nebula\cite{yang16}.

\subsection{Pulsar-like models.} 

There have been about a dozen pulsar emission models discussed in the literature, all of which suffer from criticisms either theoretically or observationally\cite{melrose17}. The only common ingredient of these many models is that electron-positron pairs, whose number density is $n_\pm = \xi n_{\rm GJ}$, are invoked. Here $n_{\rm GJ}$
is the so-called Goldreich-Julian density defined by a force-free co-rotating magnetosphere\cite{goldreich69} and $\xi \gg 1$ is the pair-production multiplicity parameter due to $\gamma$-rays interacting with strong magnetic fields or low-energy photons.

One type of pulsar coherent emission models invokes plasma effects (for example, relativistic plasma emission or certain plasma instabilities) to interpret pulsar radio emission\cite{melrose17}. These models are intrinsically plasma maser models. When applied to FRBs, it is found that none of these pulsar models can reproduce the much higher brightness temperatures of FRBs compared with pulsars\cite{lu18}. A plausible  scenario proposed recently\cite{lyubarsky20} is that a magnetic pulse generated from the inner magnetosphere of a magnetar propagates to the outer magnetosphere and triggers magnetic reconnection, which generates high-frequency fast magnetosonic waves that eventually escape as electromagnetic waves.
 
Another type of models invokes coherent curvature radiation by bunches of charged particles\cite{ruderman75}, which requires electrons/positrons to cluster in both position and momentum spaces and radiate in phase. The mechanism was invoked to interpret the general pulsar phenomenology\cite{ruderman75}, but suffers from the criticism that bunches are quickly dispersed and cannot maintain coherent emission in the observed timescale\cite{melrose78}. One suggestion to overcome this difficulty is the formation of plasma ``solitons'' due to a plasma two stream instability in the polar cap sparking model\cite{melikidze00}. This requires a highly non-stationary environment, which probably applies to the violently explosive environment of FRBs. Since the observed FRB duration is much longer than the bunch disperse timescale, this model requires a continuous generation of sparks throughout the duration of each FRB. The application of this mechanism to FRBs has been discussed by various authors\cite{katz14,kumar17,yangzhang18,lu18,lu20,yang20c}. One critical requirement is that there should be a strong electric field parallel to the local magnetic field ($E_\parallel$) in the FRB emission region\cite{kumar17}. This is because the cooling timescale of the bunches is extremely short owing to the enormous emission power of the bunches. An $E_\parallel$ would continuously supply energy to the bunches so that they can radiate for a long enough time to power the FRB. 
It is an open question how such an $E_\parallel$ could be generated. One possibility is that a linear Alfv\'en wave propagating from the surface to the outer magnetosphere would develop charge deficit and an $E_\parallel$ after reaching a critical radius\cite{kumar20a}. Such a model can account for a wide range of FRB luminosities\cite{lu20}. Within the bunching curvature radiation model, as the bunches stream out along the open field line region of a neutron star\cite{kumar17,zhang17}, the characteristic frequency of curvature radiation normally drops. If the line of sight of an observer sweeps the discrete ``spark-like'' bunches in adjacent field lines, as invoked in radio pulsar modeling\cite{ruderman75}, softer emission always occurs at later times. This provides an interpretation\cite{wang19} to the sub-pulse down-drifting feature of some bursts\cite{hessels19,chime-2ndrepeater,chime-repeaters}.

\subsection{GRB-like models.} 

The ideas of producing coherent radio bursts using relativistic shocks through the synchrotron maser mechanism has been discussed within the context of GRBs long before the discovery of FRBs\cite{usov00,sagiv02}. So far, no FRB-like signal has been detected in association with GRBs. In any case, the idea gained renewed interest in the FRB era\cite{lyubarsky14,waxman17,beloborodov17,plotnikov19,metzger19,beloborodov20}. These models invoke either internal shocks or external shocks similar to those introduced to explain prompt emission and afterglow of GRBs. One general requirement for FRB synchrotron maser models is that the bulk Lorentz factor $\gamma$ of the outflow must be much greater than unity (which means that the outflow must travel with a speed very close to the speed of light). This is because at the shock radius, the characteristic light crossing time $R_{sh}/c$ is  much longer than the duration of an FRB. A relativistic propagation effect is needed to reduce the observed timescale to milliseconds or shorter, that is, $t_{\rm FRB} \sim R_{sh}/\gamma^2 c \simeq (3 \ {\rm ms}) \ (R_{sh}/10^{12} \ {\rm cm}) / (\gamma/100)^{2}$. The size of the engine that powers the jet is of the order $R_{sh} / \gamma^2$, which is still limited by $cw$. Therefore, these models also need to invoke a compact object (for example, a magnetar in most models) as the FRB central engine.

There are several versions of synchrotron maser models. The most straightforward one is a vacuum model that neglects the plasma effect\cite{ghisellini17}. To reach negative absorption for synchrotron radiation, the field lines must be highly ordered and all electrons must have the same pitch angle. Negative absorption is achieved in the direction outside the particle  $1/\gamma_e$ emission cone (where $\gamma_e$ is the Lorentz factor of the particles). The physical conditions of this scenario are contrived.

More realistic synchrotron maser models invoke plasma effects and have two different versions. The first version invokes a weakly magnetized ($\omega_p > \omega_B$) plasma, with negative absorption occurring below the generalized Razin plasma frequency
where the electron energy distribution substantially hardens\cite{waxman17}. The maser condition is satisfied in a narrow frequency range, which can be outside the radio window where FRBs are detected. The production of GHz radiation favours a condition that the source is not much magnetized\cite{long18}.  
The second version invokes a highly magnetized ejecta with an ordered magnetic field in the pre-shock medium\cite{lyubarsky14,plotnikov19,metzger19,beloborodov17,beloborodov20}. This ordered field is required to allow shock-accelerated electrons to gyrate around the magnetic field lines and emit in phase. To achieve and maintain such an ordered field, the magnetization parameter $\sigma$ (ratio of magnetic field energy density and mass energy density in the comoving frame) cannot be much below unity. Numerical simulations\cite{plotnikov19} suggest that the efficiency of synchrotron maser in this scenario is about $7\times 10^{-4} \sigma^{-2}$ with $\sigma \gtrsim 1$. This suggests that the mechanism is very inefficient, with a large amount of ``wasted'' energy going to other channels to power emission in high-energy bands. As a shock propagates in a medium in one burst, the maser frequency decreases with time because the plasma density progressively decreases with radius\cite{metzger19,beloborodov20}. This offers a possible mechanism to interpret the frequency down-drifting in some FRBs\cite{hessels19,chime-2ndrepeater,chime-repeaters}, even though the formation of sub-pulses is not expected. In active repeaters, the same sub-pulse drifting pattern has been observed from bursts that are separated by months or even years. This requires that the shock conditions must be very similar to each other in these very different epochs. A comparison of the features of the pulsar-like and GRB-like models are presented in Box 1.

\begin{tcolorbox}
{\bf Box 1. Comparison of pulsar-like models and GRB-like models} \\
{\small The two models have the following different features that might be used to differentiate between them. In my opinion, the available data favour the pulsar-like models over the GRB-like models. \\
{\em Beaming factor.} The pulsar-like model invokes magnetic field lines to define the radiation beam, which is naturally collimated. The exact value of the beaming factor $f_b$ for FRBs is not constrained from the data. For pulsars, this can range from about $10^{-1}$ for millisecond pulsars to about $10^{-2}$ for normal pulsars\cite{lorimer12}. In the extreme case, where the beaming angle is defined by the electron Lorentz factor $\gamma_e \simeq 10^2-10^3$, $f_b$ can be as small as $10^{-5}-10^{-6}$. The total energy of each burst is (much) smaller than its isotropic equivalent value by a factor of $f_b$, and the total number of emitters (including those not seen) should be greater by a factor of $1/f_b$. The GRB-like models invoke relativistic shocks, with the geometric opening angle much greater than the $1/\gamma$ cone. Lacking a dense ambient medium (as is the case for long and short GRBs) to collimate it, the relativistic outflow (likely to be highly magnetized) launched from the engine would have a large beaming factor ($f_b \lesssim 1$). \\
{\em Efficiency:} The pulsar-like models have relatively high efficiency in producing radio emission. Observationally, the radio-to-X($\gamma$)-ray luminosity ratio for radio pulsars is typically above $10^{-4}$. Radio pulses can be continuously generated without the requirement of a long waiting time. The predicted high-energy ($\gamma$-ray/X-ray/optical) emission is moderately bright. The GRB-like models have very low radio emission efficiency. The models in general require a huge energy budget and predict a brighter high-energy counterpart than the pulsar-like models. Also a relatively long waiting time (e.g. $\gtrsim 100$ s) is needed to produce another FRB\cite{metzger19}. \\
{\em Polarization properties:} The emission in the pulsar-like models is highly polarized\cite{lu19}, which can be up to nearly 100\% linear polarization, as observed\cite{michilli18,cho20,luo20,day20}. It can account for either a constant polarization angle\cite{michilli18} (for emission from the outer magnetosphere or when the pulsar spins slowly) or a rapid swing of polarization angle\cite{cho20,luo20} (for emission from the inner magnetosphere or when the pulsar spin rapidly) during a burst. The GRB-like models have different predictions for the two sub-types: The low-$B$ version\cite{waxman17,long18} does not predict emission with strong linear polarization. The high-$B$ version\cite{lyubarsky14,beloborodov17,metzger19,beloborodov20,plotnikov19} predicts nearly 100\% linear polarization with a constant polarization angle across the pulse, but would have difficulty to account for rapid swings of polarization angles within a single burst\cite{cho20,luo20}, especially the diverse polarization angle swings among different bursts from the same source\cite{luo20}.}
\end{tcolorbox}

%

\section{Magnetars as FRB engine.}

Magnetars are neutron stars with strong magnetic fields, typically with a surface dipolar magnetic field in excess of $10^{14}$ G\cite{kaspi17}. There are 29 known magnetars in the Galaxy (http://www.physics.mcgill.ca/$\sim$pulsar/magnetar/main.html), which manifest themselves as SGRs or anomalous X-ray pulsars. They are mainly powered by the dissipation of strong magnetic fields rather than spindown. It has been hypothesized that extreme cosmic explosions such as long GRBs, superluminous supernovae, and even short GRBs due to NS-NS mergers may give birth to extremely rapidly rotating magnetars\cite{thompson93}. Galactic magnetars likely did not go through these extreme channels, since their birth rate is much higher than the birth rate of those explosions\cite{beniamini19} and since the supernova remnants of some Galactic magnetars have a moderate amount of energy inconsistent with that of a rapid rotator at birth\cite{vink06}.

Both types of magnetars have been proposed as likely sources of repeating FRBs. Shortly after the discovery of the Lorimer burst\cite{lorimer07}, SGR giant flares from regular magnetars were invoked to interpret FRBs\cite{popov10}. 
Despite the tight radio upper limit on the giant flare of SGR 1806-20 (ref. \cite{tendulkar16}), this suggestion has been put forward over the years\cite{lyubarsky14,kulkarni14,katz16,wadiasingh20}. The special host galaxy properties of FRB 121102 (ref. \cite{tendulkar17}) drove the idea that young magnetars born in long GRBs and superluminous supernovae
are the sources of repeating FRBs\cite{metzger17,beloborodov17,metzger19,beloborodov20}. Searches for FRBs in GRB or superluminous supernova remnants have been carried out and so far fruitless\cite{men19,law19} (but see ref. \cite{wangxg20}). Later observations suggest that the host galaxies of the majority of FRBs are unlike that of FRB 121102 (refs. \cite{bannister19,ravi19,marcote20,macquart20,li19}) casting doubt on the ansatz that magnetars from extreme explosions make the main sources of repeating FRBs\cite{lizhang20}.
The association between the Galactic FRB 200428 (refs. \cite{CHIME-SGR,STARE2-SGR}) and a hard X-ray burst from SGR 1935+2154 (refs. \cite{HXMT-SGR,Konus-SGR,Integral-SGR}) firmly established an FRB-magnetar connection.
The association itself and the rarity of the association\cite{lin20} suggest the following about the FRB mechanism. (1) At least some, perhaps most, even all, FRBs are produced by magnetars. (2) Magnetic energy rather than spindown energy provides the energy source of the FRB (this conclusion is based on two facts: that the FRB luminosity greatly exceeded the spindown luminosity of the magnetar and that the FRB was associated with an SGR burst believed to be triggered by magnetic dissipation event in the magnetar\cite{thompson95}); 3. Making a detectable FRB from SGR bursts requires contrived conditions. The successful rate is at most one out of a hundred.

The debate about the site and the mechanism to generate FRBs continues in light of the detection of FRB 200428. The additional information from the X-ray counterpart offers more clues, enabling us to differentiate between the models. The X-rays are about four orders of magnitude more energetic than the radio emission. This radio efficiency can be accommodated within both the pulsar-like\cite{lu20} and GRB-like\cite{margalit20} models (the latter model predicted this efficiency). There exist two hard peaks in the lightcurve of the X-ray burst that have a separation of about $30$ ms (refs. \cite{HXMT-SGR,Konus-SGR}), which coincides with the 30-ms separation between the two pulses of FRB 200428. The X-ray burst duration is much longer than the durations of the two FRB pulses\cite{HXMT-SGR,Konus-SGR,Integral-SGR}. 
It is well established that the X-ray emission originates from the magnetosphere of a magnetar\cite{thompson95}. Therefore the pulsar-like mechanism is favoured for FRB 200428, by Occam's Razor\cite{lu20,yang20c,katz20c}. A GRB-like model interprets the observed X-rays
as the superposition of magnetospheric and shock emissions, with the latter component consistent with the relatively hard spectra during the two lightcurve peaks\cite{margalit20}. Special conditions are required to satisfy the observational constraints in the shock model\cite{lu20,yu20}. The rarity of FRB-SGR associations is most probbaly due to the much narrower emission beams of the FRB emitters\cite{lin20}, which is expected in the pulsar-like models. 
Narrow FRB spectra could be another possibility, even though one has to assume that the majority of the low-frequency counterparts of SGR bursts have peak frequencies far away from the GHz band (for example, $>$ 30 GHz; ref. \cite{lin20}). Narrow FRB spectra could be produced in both the synchrotron maser (GRB-like) model\cite{waxman17,plotnikov19} and the bunched curvature radiation (pulsar-like) model with charge separation\cite{yang20c}. Finally, it is possible that the FRB coherent mechanism is delicate and only achievable in a small fraction of SGR bursts. It is also possible that all three factors play a part in causing the rarity of SGR-associated FRBs.

\begin{figure}
\begin{tabular}{c}
\includegraphics[keepaspectratio, clip, width=1\textwidth]{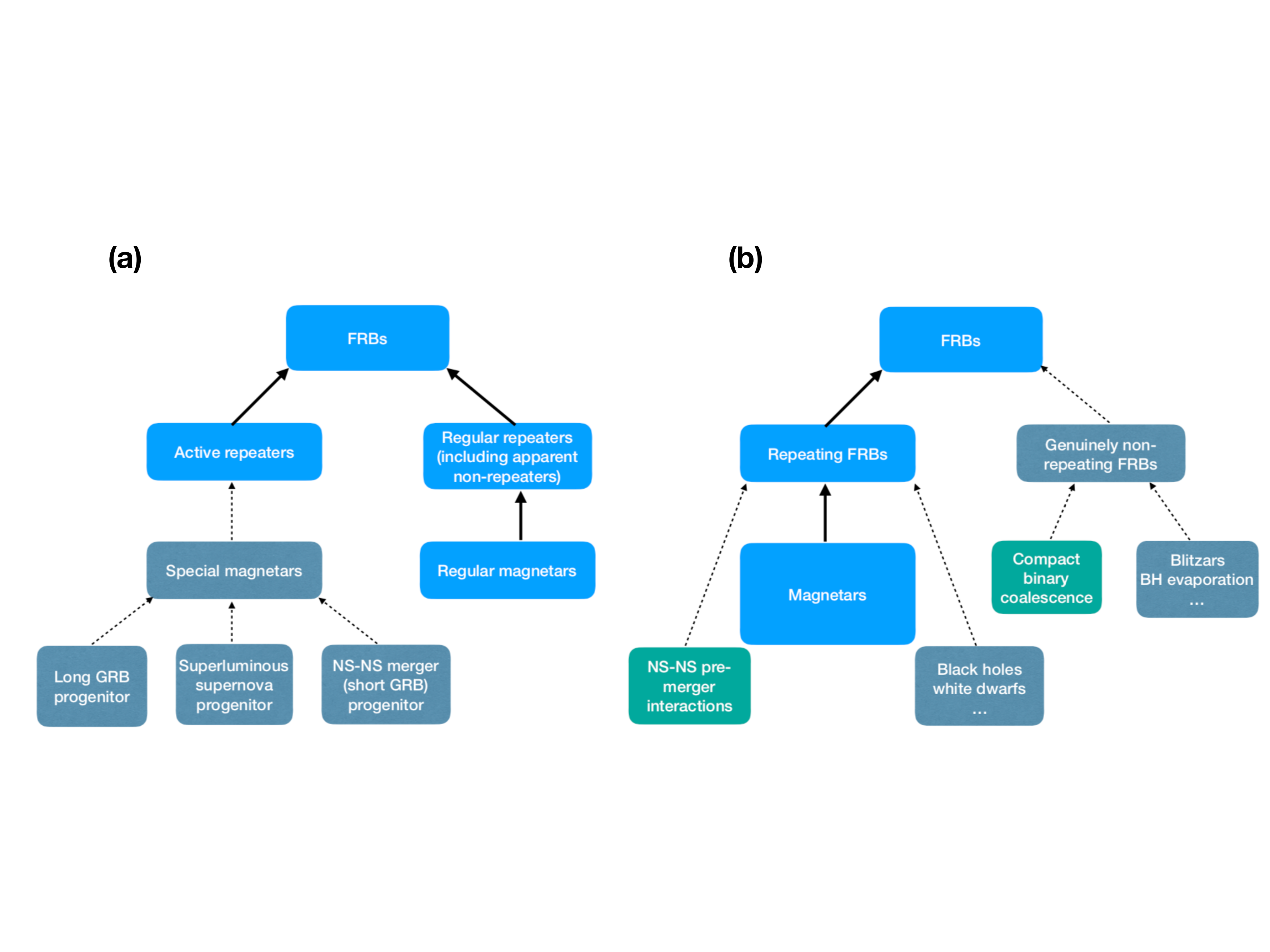}\\
\end{tabular}
\caption{{\bf Two extreme versions of FRB source models.} {\bf a,}
The most conservative scenario of FRBs, invoking a unified magnetar engine model to account for all FRBs in the universe\cite{lu20}. The light blue boxes refer to observational facts, whereas the dark blue boxes refer to speculations. The current data are not inconsistent with such a picture.
{\bf  b,} The most speculative scenario of FRBs invoking other possible engines. The light blue boxes refer to observational facts, whereas the dark blue boxes refer to speculations. The green boxes highlight the speculated multi-messemger connections. The magnetar path is the only  channel that can produce FRBs at present.}
\label{fig:magnetar}
\end{figure}

One natural question is whether magnetars can account for {\em all} the FRBs observed in the universe. The bursting rate of SGRs in the universe is about $10^2-10^3$ times more often than the FRB rate\cite{lin20,lu20}. Considering the approximately $10^{-2}-10^{-3}$ rarity factor of SGR-FRB associations, one is tempted to draw the conclusion that all (or at least the majority of) FRBs can be produced by SGR flares (Fig.\ref{fig:magnetar}a)\cite{lu20}. The fact that FRB 200428-like events fall on the extrapolation of the luminosity function of known FRBs also supports this possibility\cite{lu20}. One issue is that the number of the very active repeating FRB sources in the sky is too small to be consistent with the number of magnetars in the universe. This may require the existence of two populations of magnetars in a unified magnetar model for FRBs\cite{lu20}: a large population of regular magnetars (similar to Galactic magnetars) that contribute to the bulk of FRBs and a small population of special magnetars (those born from extreme explosions such as GRBs and superluminous supernovae) that power active repeaters. The FRB host galaxy data seem to be consistent with such a picture\cite{lizhang20}.

\section{Open questions}

Despite rapid progress in the field, there are still several open questions regarding the origin of FRBs, which
will drive the observational efforts and theoretical investigations in the field in the years to come: \\
 (1) Are there genuinely non-repeating FRBs? If so, what could be the plausible source(s)? \\
 (2) Are there engines other than magnetars that could power repeating FRBs? If so, what could be the plausible sources? \\
 (3) How is FRB emission generated, from magnetospheres (pulsar-like mechanism) or relativistic shocks (GRB-like mechanisms)? What is the correct mechanism to produce coherent emission from FRBs?

None of the three sets of questions is easy to address. Here we discuss the three sets of questions in reverse order. The first half of the third  question may find an answer as more data are collected (for differentiation criteria, see Box 1). It is not impossible that both pulsar-like and GRB-like mechanisms could operate, but again Occam's razor prefers one single mechanism given that observationally there is no obvious dichotomy in the properties of the FRB emission. Suppose supposing that the pulsar-like vs. GRB-like model debate could be settled, the prospects to answer the second half of the third question are not good. If history can be a guide, the radio pulsar coherent radiation mechanism has still not been identified after more than a half-century of study since its discovery\cite{melrose17}.

Athough it is difficult to address the second set of question by observing cosmological repeating FRBs, the detection of FRB 200428 means that we may hope to eventually answer this question with more detections of Galactic FRBs using wide-field antennas (CHIME, STARE2, and SKA) that are suitable for the detection of such bright events. The event rate density of repeating FRBs is high enough that their less energetic brethren (such as FRB 200428) can occur in the Galaxy within the timescale of interest to astronomers. For example, the detection of Galactic FRBs from young pulsars, X-ray binaries, white dwarfs or the Galactic centre would support the models that invoke those engines\cite{connor16,cordes16,katz20,gu16,zhang18b}. 
Long-term monitoring of scattered FRBs from the Moon would place robust constraints on the event rate of these events\cite{katz20b}. One limitation of this approach is that the Milky Way lacks an active repeater similar to FRB 121102. 
Current or future more sensitive $\gamma$-ray detectors may be able to detect the coincident high-energy emission associated with nearby active repeaters, establishing their magnetar origin. 
Multi-messenger observations using space-borne gravitational wave detectors (such as the Laser Interferometer Space Antenna) can help to test whether the magnetospheric interactions of binary neutron stars decades to centuries before their mergers can create an active repeating FRB source\cite{zhang20}. Most other scenarios (for example, the asteroid-neutron-star collision model\cite{dai16,smallwood19,dai20b})  
are not easy to prove or rule out unless a smoking-gun criterion can be established and tested.

The first set of question is the most difficult one to answer. Genuinely non-repeating FRBs, if any, must be only a small fraction of the detected FRB population because the observed FRB rate density greatly exceeds the event rate densities of known catastrophic events (including the beaming correction would worsen the case)\cite{ravi19b,luo20b}. Yet, some scenarios for producing short-duration FRB-like transients, such as those invoking compact binary coalescence\cite{totani13,zhang14,wang16,zhang16a,levin18} or collapse of a supra-massive neutron star (so-called ``blitzars'')\cite{falcke14}, are physically quite plausible to make short duration transients such as FRBs. Observationally, it is extremely difficult to judge whether a burst is an intrinsically one-off event, or a repeater with a very long re-activation time, even though 
long-term systematic monitoring of the same sky area (for example, the observational mode of CHIME) may offer some constraints\cite{ai20}.
Multi-messenger observations  combining the FRB search data and gravitational wave data offer an opportunity to test the FRB models related compact binary coalescence. A joint detection of an FRB and a compact binary coalescence event would unambiguously establish a category of genuinely catastrophic FRBs. Non-detection can also place progressively stringent constraints on such scenarios\cite{wangmh20}. 

In contrast to the most conservative unified magnetar model (Fig.\ref{fig:magnetar}a), Figure \ref{fig:magnetar}b displays the most speculative scenario of FRBs in which multiple astrophysical channels can produce sub-types of FRBs.
In both repeating and genuinely non-repeating categories, the models related to compact object mergers are highlighted, not only because they are attractive scenarios, but also because they can be observational tested with multi-messenger observations. 
It may be too optimistic to imagine that FRBs can form naturally in so many different ways. On the other hand, it may underestimate the richness of the universe that only magnetars can make FRBs. Future observations will bring clarity.


\begin{addendum}

 \item The author acknowledges helpful comments from two referees and the following individuals: Pawan Kumar, Wenbin Lu, Jonathan Katz. Yuan-Pei Yang, and Zi-Gao Dai.

 \item[Author Contributions]  The author conceived and wrote the review article.

 \item[Author Information] Reprints and permissions information is available 
 at www.nature.com/reprints. The author declares no competing financial 
 interests. Correspondence and requests for materials should be addressed to 
 BZ (zhang@physics.unlv.edu).

\item[Data Availability] The data that support the plots within this paper 
and other finding of this study are
available from the corresponding authors upon reasonable request.

 \item[Competing Interests] The authors declare that they have no
competing financial interests.

\end{addendum}


\end{document}